\documentclass[12pt]{iopart}
\expandafter\let\csname equation*\endcsname\relax
\expandafter\let\csname endequation*\endcsname\relax 
\usepackage{amssymb,amsfonts, amsmath}
\usepackage{algorithmic}
\usepackage{graphicx}
\usepackage{textcomp}
\usepackage{xcolor}
\newcommand{\rb}{\right)}
\newcommand{\lb}{\left(}

\begin{document}

\title[Reducing hyperparameter dependence by external timescale tailoring]{Reducing reservoir computer hyperparameter dependence by external timescale tailoring
}

\author{Lina Jaurigue and Kathy Lüdge}

\address{Technische Universität Ilmenau}
\ead{lina.jaurigue@tu-ilmenau.de}
\vspace{10pt}

\begin{abstract}

Task specific hyperparameter tuning in reservoir computing is an open issue, and is of particular relevance for hardware implemented reservoirs. We investigate the  influence of directly including externally controllable task specific timescales on the performance and hyperparameter sensitivity of reservoir computing approaches. We show that the need for hyperparameter optimisation can be reduced if timescales of the reservoir are tailored to the specific task. Our results are mainly relevant for temporal tasks requiring memory of past inputs, for example chaotic timeseries prediciton. We consider various methods of including task specific timescales in the reservoir computing approach and demonstrate the universality of our message by looking at both time-multiplexed and spatially multiplexed reservoir computing.

\end{abstract}

\section{Introduction}

Reservoir computing is a machine learning method that has received a lot of research attention recently. There are several reasons for this. Firstly, the learning method is very simple; only the output layer is trained via linear regression \cite{JAE01,MAA02}. This makes the implementation of the reservoir computing relatively straightforward and therefore accessible for researcher form a wide range of backgrounds. Secondly, reservoir computing has been shown to perform well on predicting complex dynamics \cite{NAK21,BUE17,KUR18,SHA22}. And, thirdly, due to the simplicity of the training method it is well suited for hardware implementation \cite{SAN17a,TAN19c}. In particular, time-multiplexed reservoir computing of delay-based reservoirs has been a popular method of hardware implementation.
The idea to utilize a feedback delay loop instead of a network was first published in \cite{APP11} and has since then been implemented and tested for opto-electronic \cite{APP11,LAR12,PAQ12,CHE19c}, optical \cite{BRU13a,DEJ14,VIN15,HOU18, BOR21,DON22} and electromechanical \cite{DIO18} realizations. We also refer the reader to recent reviews on this topic \cite{BRU18a, BRU19, NAK20, TAN19c, NAK21,CUC22}.

An issue that remains is the lack of an efficient method for optimising the parameters (hyperparameters) of the reservoir in a task specific manner. This is particularly relevant for hardware implemented reservoir computing, where parameters can be difficult to tune. There have been many studies that have looked into various approaches including grid searches, Bayesian optimisation, genetic algorithms, or new network measures that correlate well with the reservoir performance \cite{PEN18,GRI19,RAC21,VAL23,CAR20a}. However, these approaches can greatly increase the computational cost of the training phase, or are not suitable for hardware implementation. In this paper we aim to contribute to solving the issue of hyperparameter optimisation, with a particular focus on hardware implementation. Specifically, we investigate the influence of augmenting the memory of the reservoir at either the input or the output level such that task relevant timescales are present in the system.

For the case of delay-based reservoirs it is known that
the feedback delay-time can be an important tuning parameter, as it directly influences the memory of the reservoir \cite{HUE22,KOE21}. In the delay-based reservoir computing community, the delay-time is mostly chosen to be either resonant with the time-scale of the input clock-cycle or offset from it by one input mask interval. However, depending on the task, other values of the delay-time are beneficial \cite{ROE19, STE20}. We therefore also address the influence of tuning the memory of the reservoir directly via the internal delay as one possible optimization step.

The fact that task relevant timescales play an important role in timeseries prediction tasks is well established and in the context of reservoir computing is also closely related to Takens embedding \cite{TAK80,HAR20b}. In this paper we highlight how we can externally add task relevant timescales by adding additional delay lines at the input or output layer. These approaches are universal for all reservoir computing realizations, including electrical, optical and digital implementations, and should not significantly decrease the speed or the energy efficiency in hardware implementation. 
We show that by externally adding the required memory to the reservoir, the performance can be improved over the hardware relevant parameter spaces, thereby reducing the need for hyperparameter optimisation.




When optimising the timescales of a reservoir for a particular task, the discretisation of the input timeseries also plays an important role.
The authors of \cite{STO22} have performed a comprehensive investigation of closed-loop timeseries prediction using ESNs. They consider the influence of the discretisation of the input timeseries and look at the difference between providing full and partial information. While, in \cite{TSU23} the authors consider the relationship between discretisation and dynamic timescales. We also discuss the influence of the discretisation here, but with a focus on the resulting memory requirements on the reservoir.






There have been several other recent studies into various methods of augmenting the reservoir computing approach.
In \cite{DEL21a} and \cite{CAR22a} the authors concatenate the state matrix with randomly selected delayed system states and the authors of \cite{PIC23} use different delayed output information for improving action recognition with a frequency multiplexed reservoir computing system. In \cite{HOL10} trainable state delays are investigated. In \cite{SAK20} and \cite{MAR19a} concatenation of the state matrix with past states is studied in the context of reservoir size reduction. 
The authors of \cite{GAU21b} highlight the possibility of forgoing the reservoir by directly filling the state matrix with nonlinear transforms of past inputs, in which case the memory can be freely chosen. This nonlinear vector regression approach, referred to as "next generation reservoir computing", can produce accurate results for standard benchmark tasks at a relatively low computational cost, however the hardware implementability is greatly reduced and the optimisation of hyperparameters is traded for the selection of inputs and nonlinear functions \cite{JAU22}.
In the context of hardware implemented delay-based reservoir computing, recent efforts toward performance enhancement include using ensembles of delay-based reservoirs \cite{ORT17a}, sequentially coupled delay-based reservoirs \cite{PEN19,GOL20} or  delay-coupled networks \cite{ROE19}. Our contribution is to highlight the importance of task specific timescales, and to show how these can easily and tunably be incorporated in hardware implemented reservoir computing.

The paper is structured as follows. The concepts of reservoir computing and the external memory augmentation approaches are introduced in Sec.\,\ref{RC} and \ref{Sec:Memory}, respectively. In Sec.\,\ref{Sec:Discret} we briefly discuss timeseries descretisation. The reservoir computing models and tasks used in this study are introduced in Sec.\,\ref{Sec:Task}. The results showing the improved performance with augmented memory and the influence of the input timeseries discretisation are presented in Sec.\,\ref{Sec:Results}, before we conclude in Sec.\ref{Sec:Conclusion}.

\section{Reservoir Computing}\label{RC}

Reservoir computing is a scheme wherein the input response of a dynamical system (the reservoir) is sampled a number of times and these sampled system responses are then linearly combined to construct an output. During the training phase, the weights of the linear combinations of the system responses (output weights) are trained via linear regression. This is done by feeding the reservoir a sequence of $K_T$ inputs and collecting $S$ responses of the reservoir to each input into a state matrix ${\bf S}\in \mathbb{R}^{K_T}\times \mathbb{R}^{S+1}$, adding a bias term of one to each row, and then finding the vector of weights ${\bf w} \in \mathbb{R}^{S+1}$ that minimises the difference between the target output ${\bf y}\in \mathbb{R}^{K_T}$ and the output ${\bf \hat{y}}={\bf 
S}{\bf w} \in \mathbb{R}^{K_T}$. The solution to this problem is given by 
\begin{equation}\label{Eq:wout}
    {\bf w}=({\bf S}^\textrm{T}{\bf S}+\lambda {\bf I})^{-1} {\bf S}^\textrm{T} {\bf y},
\end{equation}
using the Moore-Penrose pseudoinverse. Equation \eqref{Eq:wout} includes Tikhonov regularisation, with the regularisation parameter $\lambda$ and identity matrix ${\bf I} \in \mathbb{R}^{S+1}\times \mathbb{R}^{S+1}$.

\subsection{Spatially-multiplexed Reservoir Computing}

The responses of the reservoir can either be multiplexed in space or in time (or combinations whereof). Echo-state networks are an example of spatially multiplexed reservoirs \cite{JAE01}. In this scenario, there are a number of spatially separated dynamical nodes which evolve discretely in time. These nodes are fed with a weighted input signal and the responses of some (or all) of the nodes are collected into the state matrix. The evolution of such a reservoir can be described by an equation of the form
\begin{equation}\label{eq:ESN}
    {\bf x}\lb k+1\rb=f\lb {\bf W}_{int}{\bf x}\lb k\rb +{\bf w}_{in}I\lb k +1\rb\rb,
\end{equation}
where ${\bf x}\lb k\rb\in \mathbb{R}^M$ is the vector describing the state of the $M$ nodes at time $k$, ${\bf W}_{int}\in \mathbb{R}^{M}\times \mathbb{R}^{M}$ is the matrix of internal reservoir node couplings (adjacency matrix), $I\lb k\rb$ is the input signal at time $k$, and ${\bf w}_{in}\in \mathbb{R}^M$ is the vector of input weights. The function $f \lb \cdot\rb$ is a nonlinear function which is applied element wise and commonly referred to as the activation function. In Eq.~\eqref{eq:ESN} the input is one dimensional, i.e. there is a single temporal input signal. To generalise to $p$ input signals, $I\lb k \rb$ becomes a vector ${\bf i}_{in}\lb k\rb \in \mathbb{R}^p$ and the vector of input weigths ${\bf w}_{in}$ becomes a matrix ${\bf W}_{in}\in \mathbb{R}^M\times \mathbb{R}^{p}$.

The $k^{\textrm{th}}$ row of the state matrix is filled by sampling $S$ of the reservoir states ${\bf x}\lb k\rb$, with $S \leq M$. For $S=M$ the elements of the state matrix are $s_{k,j}=x_j\lb k\rb$ with $x_j\lb k\rb$ ($j\in \left[0,..S\right)$) being the elements of the reservoir state vector ${\bf x}\lb k\rb$.

Spatially continuous reservoirs are also possible. In such a case the reservoir evolution would be described by a suitable system of partial differential equations, and the state matrix would be filled by sampling discretely in space (and in time).

\subsection{Time-multiplexed Reservoir Computing}
In the case of time-multiplexed reservoir computing, typically, a masked input signal is sequentially fed into the reservoir and the $S$ reservoir responses needed to fill the state matrix are sampled sequentially in time. The input mask is analogous to the input weights in the spatially multiplexed case, but rather than simultaneously feeding the input signal into spatially separated nodes with varying weights, the input signal is fed into the same node multiple times with varying weights (mask values).  
If $I\lb k\rb$ is the task-dependent input signal at time $k$, then the masked reservoir input is given by 
\begin{equation}\label{eqmask1}
J\lb i \rb=J\lb k,j \rb=I\lb k \rb m_1 \lb j \rb, 
\end{equation}
where $i=Sk+j$, $k\in \left[0,..K_T\right)$, $j\in \left[0,...,S\right)$ and $m_1\lb j\rb$ are the mask values. 
The state matrix entry corresponding to the input $J\lb i\rb$ is given by $s_{k,j}=x\lb i\rb=x\lb k,j\rb$, where $x\lb i\rb$ is the sampled reservoir response. If the reservoir is a time-continuous system, then the discrete input sequence can be transformed into a piecewise constant function $\tilde{J}\lb t\rb$ by holding each input constant for a fixed time $\theta$: In such a case the time for one input mask  sequence, $\theta S$, is typically referred to as the clockcycle $T$ \cite{HUE22}. In this study we consider a time discrete reservoir, see Sec.~\ref{Sec:model}, in which case the dimensionless clockcycle $\tilde{T}$ is equal to the number of input mask values $S$.

Here we have taken the number of mask values to be equal to the number of sampled reservoir responses $S$. This is not a necessary condition, but is the typical choice. We adhere to this convention for simplicity. We also note that, in delay-based reservoir computing the number of sampled reservoir responses $S$ is often referred to as the number of virtual nodes.

\section{Reservoir Computing Memory Augmentation Methods}\label{Sec:Memory}

The memory requirements of timeseries prediction tasks are very task specific and can vary greatly, even for tasks using the same input signal but trained on different target signals, e.g. for different prediction horizons. To perform well, the reservoir must have memory on the relevant timescales. This means that there must be information of past values of the input signal in the $k^{\textrm{th}}$ row of the state matrix in order to accurately predict the $k^{\textrm{th}}$ target $y \lb k\rb$. Note that here we are using the term "memory" to refer to any influences of past inputs on the state of the reservoir, i.e. also nonlinear transforms of past inputs. In the typical reservoir computing scheme, this memory comes from the dynamics of the reservoir. In this study we compare methods of augmenting the memory of a reservoir.
There are many ways in which this can be done and the choice of method comes down to the objective one aims to achieve, which means there is always a tradeoff between computational cost and absolute performance.
The motivation of this study is to reduce the need for hyperparameter optimisation at a low computational cost, as well as allowing for easy hardware implementation. To this end, we investigate two simple schemes of augmenting the memory externally at the levels of the input and the output layer of the reservoir. These two methods are described below.

\begin{figure}[t]
\centerline{\includegraphics[width=0.95\textwidth]{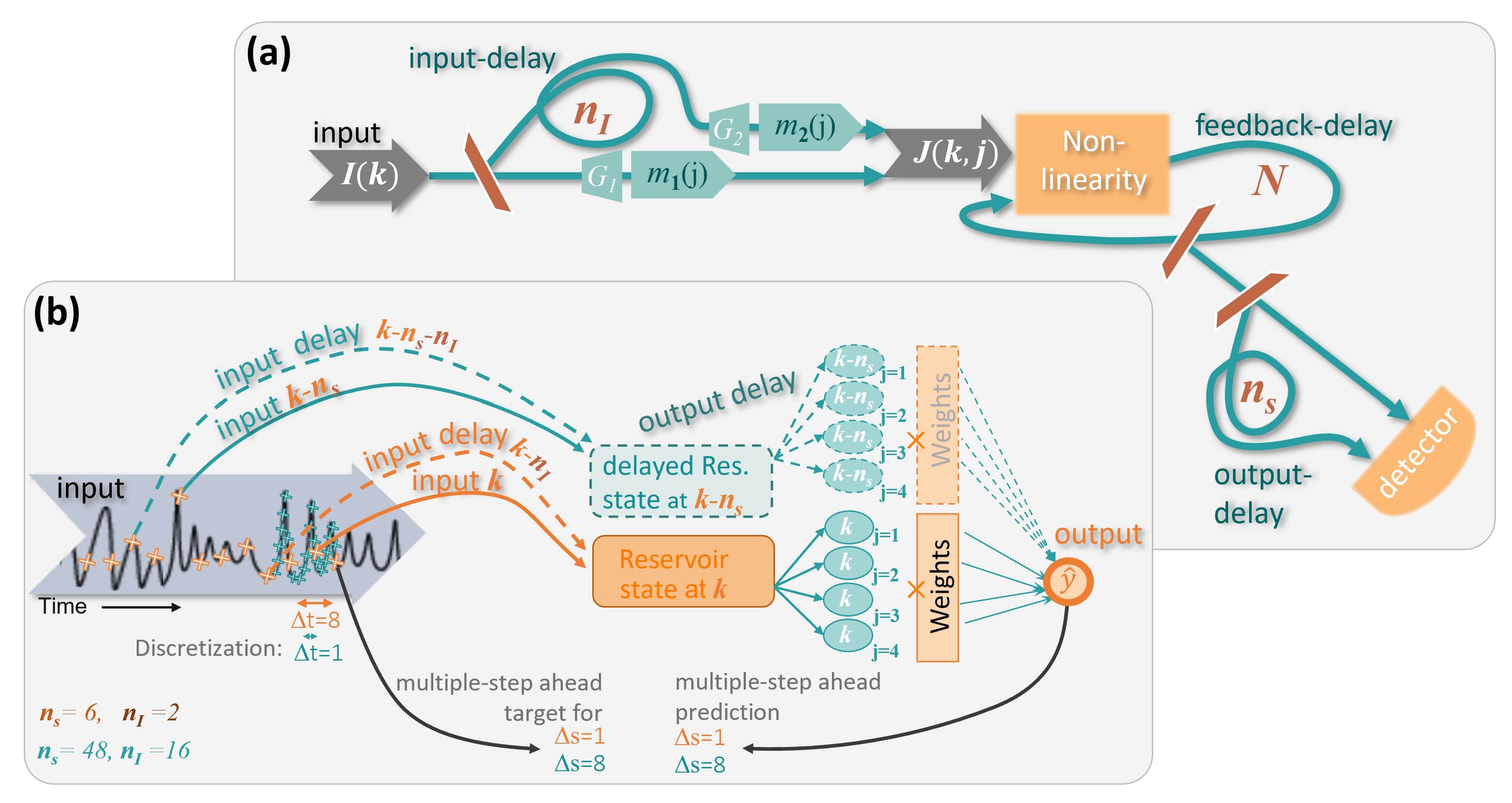}}
\caption{Sketch of (a) a possible implementation for a delay-based reservoir with additional delayed input and state matrix concatenation (delayed output), and (b) an example of the input/output data processed in these external delay schemes to predict a target (orange crosses and bended arrows indicate the situation for $n_S=6, n_I=2, \Delta s=1$).}
\label{sketch}
\end{figure}

\subsection{Delayed-input}\label{Sec:indel}

We consider the influence of one additional time-delayed input implemented as in \cite{JAU21a,JAU22x}. The input function Eq.~\eqref{eqmask1} is modified to
\begin{equation}\label{eqmask2}
\tilde{J}\lb i \rb=I\lb k \rb m_1 \lb j \rb+I\lb k-n_I \rb m_2 \lb j \rb,
\end{equation}
where $n_I$ is the input-delay and $m_2 \lb j \rb$ is the mask applied to the delayed version of the input sequence, $I\lb k-n_I \rb$. This delayed input scheme is illustrated in Fig.~\ref{sketch}b.

In terms of a photonic setup, such a delayed input could be implemented by adding a delay-loop before the reservoir, as depicted in Fig.~\ref{sketch}a. Since $n_I$ describes the input shift via the number of k-steps, the actual length of the delay line depends on the  sampling interval.

\subsection{Delayed State Matrix Concatenation (Output delay)}\label{Sec:statemat}

On the level of the output layer, a method of including a task-dependent delay, is to concatenate the current and delayed versions of the reservoir responses, i.e. to concatenate the state matrix with a delayed version of itself:
\begin{equation}
\tilde{{\bf S}} = 
\begin{pmatrix}
s_{1,1}  & \cdots & s_{1,S} & s_{1-n_S,1}  & \cdots & s_{1-n_S,S} &1\\
\vdots  & \ddots & \vdots  &\vdots    & \ddots & \vdots &1\\
s_{K_T,1}  & \cdots & s_{K_T,S} & s_{K_T-n_S,1}  & \cdots & s_{K_T-n_S,S} &1
\end{pmatrix},
\end{equation}
where $n_S$ is the output delay that described how many steps (index $k$) we go back to collect the input for the state matrix, see Fig.~\ref{sketch}b. 
This approach has recently been investigated in \cite{MAR19a} and \cite{SAK20} in the context of reducing the size of a recurrent neural network reservoir. 

In order to make fair comparisons with other methods it is important to consider whether one should compare results using the same reservoir sampling dimension $S$ (number of virtual nodes), or whether this should be halved such that the final state matrix dimensions are the same. The latter case has the advantage that the dynamical system can be smaller, or that the input clockcycle can be reduced in the case of time multiplexed reservoir computing. In this study we choose to keep the reservoir sampling dimension $S$ the same because in the context of hardware implemented reservoirs we believe comparing the same sized physical reservoirs is fairer.

In a photonically implemented reservoir, this approach could be incorporated by including an additional delay-loop after the reservoir, as sketched in Fig.~\ref{sketch}a. Again,  the physical length of the delay line is given by the product of $n_S$ and the sampling time.

\section{Input Discretisation}\label{Sec:Discret}

Another approach to improve the reservoir performance by incorporating task specific memory requirements is to modify the task itself. When constructing a time series model from data, the discretisation $\Delta t$ (sampling step) of the input timeseries can have a significant influence on the model and on the resulting performance \cite{KAN03, TSU23}. In the context of reservoir computing, the input discretisation influences the memory required of the reservoir and  is related to Takens (delay) embedding \cite{TAK80}. Recent studies have shown that under certain conditions reservoirs perform a delay embedding of the input timeseries \cite{HAR20b,DUA23}. For this to occur, and to yield accurate results, the reservoir must have memory on the appropriate time scales, which are related to the dynamics of the task \cite{FRA86}. This means that if the discretisation of the input timeseries is finer than is necessary, the reservoir will need to be able to remember input steps which lie further in the past in order for it to perform the delay embedding. The impact of different $\Delta t$ is discussed in Sec.\,\ref{Sec:Resultsdiscretisation}.



\section{Tasks and Methods}\label{Sec:Task}

\subsection{Reservoir models}

We investigate the influence of the memory augmentation methods focusing mainly on a time-multiplexed delay-based reservoir. The methods are, however, not specific to such reservoirs. We therefore also show an example of a spatially multiplexed reservoir (i.e. a random recurrent network reservoir). Both reservoirs are described below.

\subsubsection{Time-delayed Reservoir Model}\label{Sec:model}
~\\
The time-delayed reservoir model we consider is the following iterative map: 
\begin{equation}\label{eq1}
x \lb i \rb=\mathcal{G}\lb Kx\lb i-N \rb+J(i)\rb,
\end{equation}
where $\mathcal{G} \lb \cdot \rb$ is the nonlinear activation function, $J\lb i \rb$ is the input sequence given by Eq.~\eqref{eqmask1}, $K$ is the strength with which the current iteration is coupled with a past state of the system and $N$ is the feedback delay (see Fig.\,\ref{sketch}a).  We remind the reader that the index $i$ runs over all mask steps and input points ($i=Sk+j$).

Motivated by photonic reservoir implementations, we use a nonlinear function which models a semiconductor optical amplifier (SOA) for the activation function. An SOA can be described by a dynamical differential equation model which describes the light-matter interaction and the complex carrier dynamics \cite{ZAJ17,LIN17}. In our case, only the static characteristic is important, which is why we use the simple input-response of an SOA that is given by \cite{COL12a}:
\begin{equation}
	\mathcal{G}\lb x\rb =\frac{g_0 x}{1+x/a}.
\end{equation}
Using this activation function, Eq.~\eqref{eq1} describes the input-response of an SOA with weak time-delayed feedback in the limit where the time between mask steps is much longer than the characteristic time-scale of the carrier dynamics in the SOA, i.e. the limit that the reservoir reaches a steady state before it is sampled at the end of each mask step. That means, we do not include any dynamic response of the SOA and all the memory of the setup comes from the delay-loop. A sketch of such a setup is depicted in Fig.~\ref{sketch}a.

For the input masks in Eqs.~\eqref{eqmask1} and \eqref{eqmask2} we use a set of $S$ randomly drawn values from a uniform distribution between zero and one, $U\lb 0,1\rb$, and multiply these by a scaling factor $G_\alpha$, i.e. $m_\alpha \lb j\rb =G_\alpha U^\alpha_j$ for $\alpha=1,2$ and $U^\alpha_j$ sampled from $U\lb 0,1\rb$, where $G_1$ is the input scaling of the current input and $G_2$ the scaling factor for the delayed input in Eq.\eqref{eqmask2}. 

Unless stated otherwise, the parameters for the time-delayed reservoir are as given in Tab.~\ref{tab1}.

\begin{table}[htbp]
\caption{Dimensionless time-delay reservoir and input parameters.}
\begin{center}
\begin{tabular}{c|c|c|c}
\hline
\multicolumn{4}{c}{\textbf{Time-delayed Reservoir Parameters}} \\
\hline
\textbf{\textit{Parameter}}& \textbf{\textit{Value}}& \textbf{\textit{Parameter}}& \textbf{\textit{Value}} \\
\hline
SOA gain parameter~~$g_0$	&40		& node number ~~$S$	& 250\\
SOA nonl. parameter ~$a$ & 1 &clock cycle~~$\tilde{T}$ & 250\\
regression parameter~~$\lambda$ & 5e-6 & input gain~~$G_1$ & 0.2\\
number training steps & 10000& internal delay ~$N$ & 251\\
number of testing steps & 5000 & feedback strength~~$K$ & 0.03\\
number of realizations~~$Z_R$ & 10 & &\\
\hline
\end{tabular}
\label{tab1}
\end{center}
\end{table}

\subsubsection{Random Network Reservoir Model} \label{Sec:modelrand}
~\\
We use a random network as described by Eq.~\eqref{eq:ESN} with 
\begin{equation}
    f\lb x\rb = \Theta\lb x-1\rb,
\end{equation}
where $\Theta \lb \cdot \rb$ is the Heaviside step function. The internal node coupling weights are given by ${\bf W}_{int}=K_r{\bf W}_{int}^u$ where the entries of ${\bf W}_{int}^u$ are chosen randomly from a uniform distribution between zero and one, and $K_r$ is a scaling factor that determines and coupling strength and the spectral radius $\rho$. Analogous to the time-delayed reservoir, the input weights are also sampled from $U\lb 0,1\rb$ and scaled by a factor $G_1$.

The motivation for our choice of random network is not to achieve the best performance possible, but rather to demonstrate the universality of our results on a reservoir with very different properties to the time-delayed reservoir described above.

\subsection{Tasks}

We have chosen three standard benchmark tasks to demonstrate the influence of the memory augmentation methods; multi-step-ahead prediction of the Mackey-Glass chaotic timeseries, one-step-ahead prediction of the $X$ variable of the Lorenz 63 system \cite{LOR63} and cross prediction of the $Z$ variable of the Lorenz 63 system. Another method of characterising the nonlinear transform and memory capabilities of reservoir computers is to calculate the information processing capacity \cite{DAM12}. However, since it is difficult to relate the information processing capacities to task specific performance, particularly for tasks requiring temporally correlated inputs \cite{HUE22a}, we have chosen leave this approach for future studies. The Mackey-Glass and Lorenz tasks are described below.

\subsubsection{Mackey-Glass Time-series Prediction}\label{Sec:MG}

\begin{figure}[t]
\centerline{\includegraphics[width=0.9\textwidth]{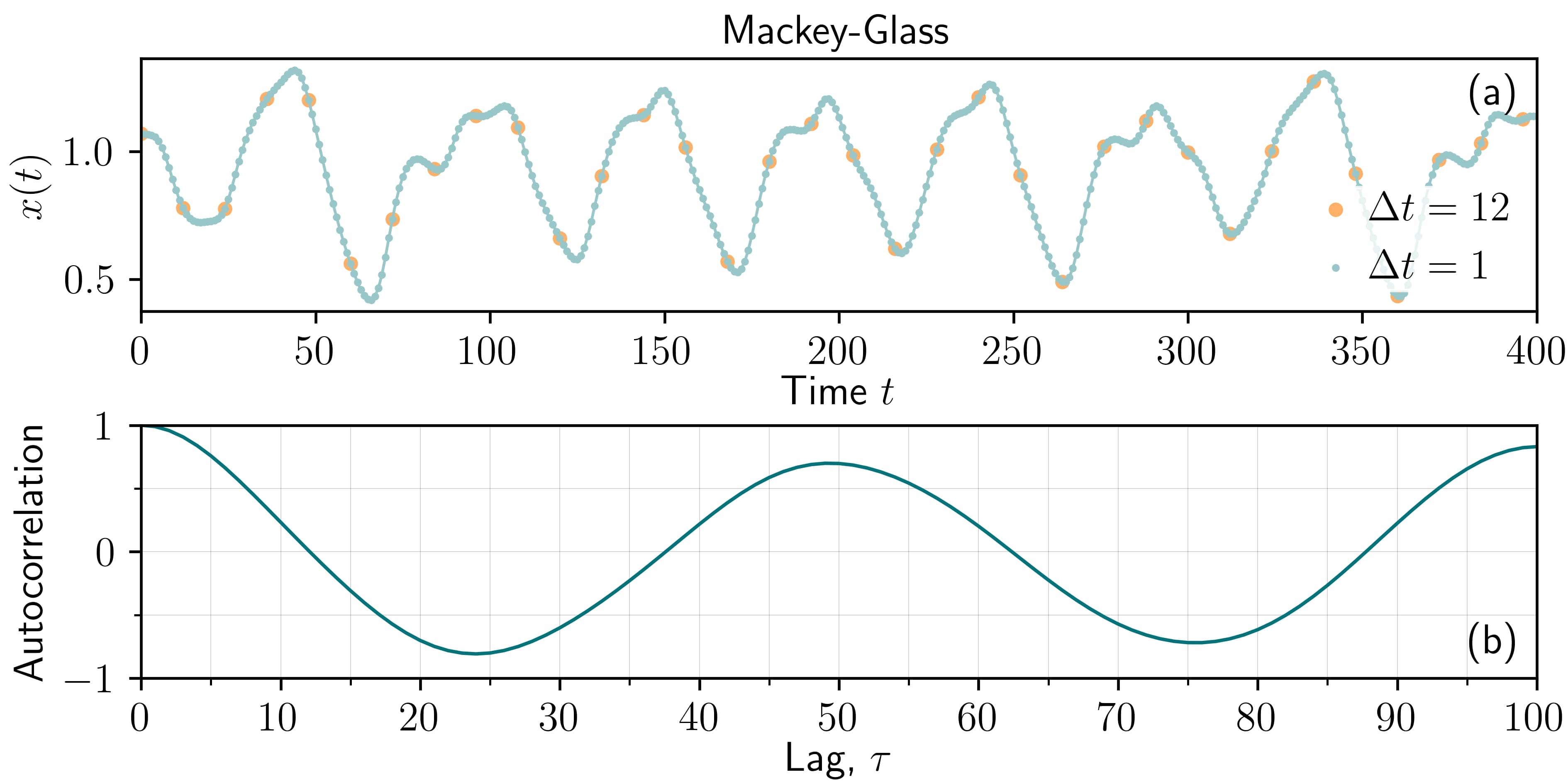}}
\caption{\textbf {Mackey-Glass task:} (a) Timeseries as given by Eq.\,\eqref{eqMG} showing chaotic dynamics for two different discretisation steps, $\Delta t=12$ (orange dots) and $\Delta t=1$ (green dots), (b) and the corresponding normalized autocorrelation $\langle x(t) x(t-\tau) \rangle$ function as a function of the time lag $\tau$.}
\label{corrMG}
\end{figure}
~\\
Multi-step-ahead prediction of timeseries produced by the Mackey-Glass equation is a common benchmarking task \cite{SHI07,ORT15,GOL20}. The Mackey-Glass delay-differential equation is given by \cite{MAC77}
\begin{equation}\label{eqMG}
\frac{dx}{dt}=\beta \frac{x \left( t-\tau_M \right) }{1+x \left( t-\tau_M \right)^n}-\gamma x.
\end{equation}
Using the standard parameters, $\tau_M=17$, $n=10$, $\beta=0.2$ and $\gamma=0.1$, this system exhibits chaotic dynamics. 

In this study we focus mainly on $\Delta s=12$ step-ahead prediction of this system with an input discretisation of $\Delta t=1$ (open-loop). 
To create the input sequence $I\lb k \rb$, we generate a timeseries by numerically integrating Eq.\eqref{eqMG} using a Runge-Kutta fourth order method with Hermitian interpolation for the mid-steps of the delayed terms and with a time step of $h=10^{-2}$. This timeseries is then sampled with a time step of $\Delta t=1$. The corresponding target sequence is given by $I\lb k+\Delta s \rb$. We have chosen $\Delta t=1$ as our default sampling, as this is a commonly used discretisation and it is small enough that finer features of the time series can be resolved, as can be seen by the green dots in Fig.\ref{corrMG}a. It is however not the optimal discretisation for delay-embedding \cite{KAN03}.

The chaotic Mackey-Glass timeseries is shown in Fig.~\ref{corrMG}a, along with its cor\-res\-pon\-ding auto-cor\-re\-la\-tion function in Fig.~\ref{corrMG}b. The dynamics are chaotic, but exhibit clear os\-cil\-la\-tions on a scale of approximately 50 time units. For the reservoir to reconstruct the dynamics of this timeseries, it would need to have memory on the order of 0.1-0.5 of these features \cite{FRA86}.
Furthermore, the autocorrelation shows that points separated by one time unit are highly correlated. This indicates that, with an input discretisation of $\Delta t=1$, consecutive inputs will contain redundant information and the reservoir will need to be able to retain information from about 10 steps into the past for a delay embedding to be performed. We therefore also consider other combinations of $\Delta t$ and $\Delta s$ to compare the influence of the discretisation of the input signal on the memory requirements of the reservoir and the performance.

\subsubsection{Lorenz 63 Time-series Prediction}\label{Sec:Lorenz}
\begin{figure}[t]
\centerline{\includegraphics[width=0.9\textwidth]{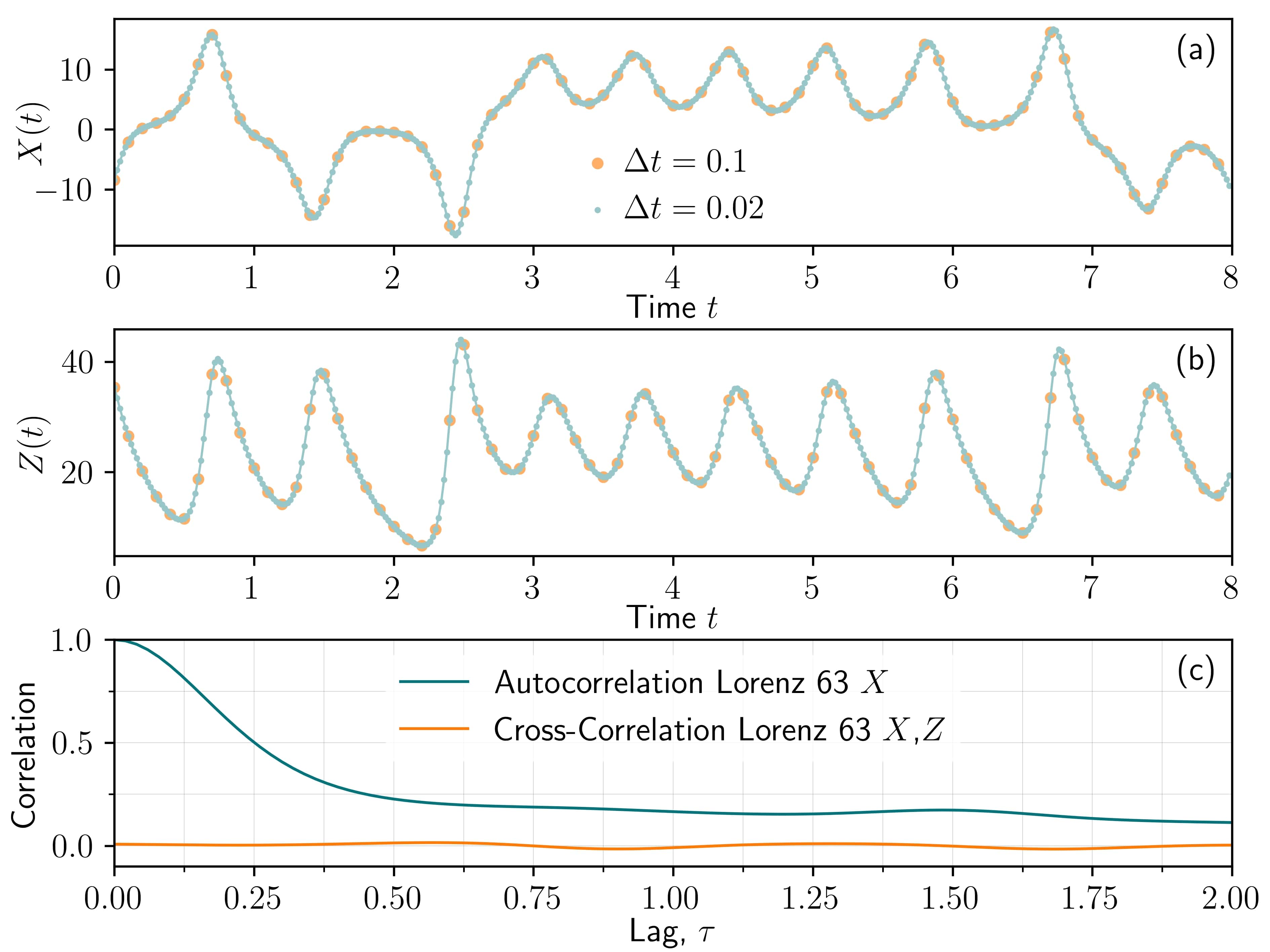}}
\caption{\textbf {Lorenz 63 task:} Timeseries for X (a) and Z (b) as given by Eq.\,\eqref{Lorenz} (chaotic dynamics) for two different discretisation steps, $\Delta t=0.1$ (orange dots) and $\Delta t=0.02$ (green dots). (c) The corresponding normalized autocorrelation $\langle X(t) X(t-\tau) \rangle$ and cross correlation function $\langle X(t) Z(t-\tau) \rangle$.}
\label{corrLor}
\end{figure}

~\\
The Lorenz system \cite{LOR63} is given by 
\begin{eqnarray}
\frac{dX}{dt}=c_1Y-c_1X,\quad
\frac{dY}{dt}=X(c_2-Z)-Y, \quad \textrm{and} \quad
\frac{dZ}{dt}=XY-c_3Z.\label{Lorenz}
\end{eqnarray}
With $c_1=10$, $c_2=28$ and $c_3=8/3$ this system exhibits chaotic dynamics as can be seen for the X and Z coordinate in Fig.\,\ref{corrLor}a and b. We use the $X$ variable, sampled with a step size of $\Delta t=0.1$ (Fig.\,\ref{corrLor} orange dots), as the input $I\lb k \rb$ for two time series prediction tasks. The first task is one step-ahead ($\Delta s=1$) prediction of the $X$ variable. The second task is cross prediction of the $Z$ variable ($\Delta s=0$). For the cross prediction task we also consider a discretisation of $\Delta t=0.02$ (Fig.\,\ref{corrLor} green dots). We generate the Lorenz timeseries using Runge-Kutta fourth order numerical integration with a time step of $h=10^{-3}$.

When doing timeseries prediction using an ordinary differential equation system like the Lorenz 63 system, there is a qualitative difference between the tasks where all dynamical variables are provided as input and tasks where this is not the case. Using the Lorenz 63 system, a comparitive study of these two cases was made in \cite{STO22}. When all variables are provided, no memory of past steps is needed to determine the evolution of the dynamics. However, if only partial information is provided, then the reservoir must perform a delay embedding \cite{HAR20b}. For our study, only the latter case is relevant as we only provide one of the three dynamical variables for the Lorenz 63 task (and not the complete history for the Mackey-Glass task). The sampling step for an optimal embedding for the Lorenz 63 system is about 0.1 \cite{KAN03,TSU23}. Figure~\ref{corrLor}c shows the autocorrelation of the $X$ variable (green line) and the cross-correlation between the $X$ and $Z$ variables (orange line) for the chaotic dynamics (Fig.~\ref{corrLor}a,b) with the parameters given above. We will see later on in Sec.~\ref{Sec:Lresults} that the differences in the correlation between the input and the target for the $X\lb t \rb \rightarrow X\lb t +\Delta t \rb$ and $X\lb t \rb \rightarrow Z\lb t  \rb$ tasks have important consequences for the optimal discretisation.

\subsection{Error Measure}

We quantify the reservoir computing performance using the normalised root mean squared error (NRMSE), defined as
\begin{equation}
 \textrm{NRMSE}=\sqrt{\frac{\sum_{k=1}^{K_T}\lb y_{k}-\hat{y}_{k}\rb^2}{K_T \textrm{var}\lb {\bf y}\rb}},
\end{equation}
where $y_k$ are the target values, $\hat{y}_{k}$ are the outputs produced by the reservoir computer, $K_T$ is the number of testing steps (i.e. the length of the vector ${\bf y}$) and $\textrm{var}\lb {\bf y} \rb$ is the variance of the target sequence.

\subsection{Simulation Parameters}

For all tasks we rescale the input and target sequences to the range between zero and one.
Before beginning the training and testing phases for these tasks, we initialise the reservoirs by feeding in 10000 inputs $I\lb k \rb$. In the training phase we use 10000 inputs while in the testing phase 5000 inputs are used. All results presented in the next section show the testing error averaged over $Z_R$ realisations of the random input masks (or input weights). For the NRMSE results the associated error bars are the standard deviation of the $Z_R$ realisations.

\section{Results and Discussion}\label{Sec:Results}

\subsection{Mackey-Glass task performance comparison}\label{Sec:MGTDR}

We are interested in the influence of including task specific timescales in the reservoir computing scheme. The delayed-input and delayed-state matrix methods described above can be applied to all reservoir computing schemes. However, when using a delay-based reservoir, we also have the option of directly tuning the memory properties by changing the feedback delay time. For the reservoir used here (Eq.~\eqref{eq1}), this corresponds to the parameter $N$.
In most delay-based reservoir computing studies, the authors typically choose $N=\tilde{T}$ (delay-time $\tau=T$ in the time-continuous case) or $N=\tilde{T}+1$ ($\tau=T+\theta$) \cite{APP11, PAQ12, BRU13a, DEJ14, TAK18, ORT17a,SUG20, ARG20}, where $\tilde{T}$ is the input clockcycle. 
It has been shown that resonances between the feedback delay time and clock-cycle are generally detrimental to the reservoir computing performance \cite{STE20,KOE21,HUE22}, we therefore choose the desynchronised configuration $N=\tilde{T}+1$ ($N=S+1$) as our default and compare its impact with the memory augmentation methods.
The other parameters of the time-delayed reservoir are the feedback strength $K$, the input scaling $G_1$ and the parameters of SOA the nonlinearity $g_0$ and $a$. The latter two are determined by properties of the SOA and are therefore not freely tunable. Since we are motivated by hardware implementability, in the following, we only consider the influence of $K$ and $G_1$, while keeping the material specific parameters $g_0$ and $a$ fixed.

\begin{figure}[t]
\centerline{\includegraphics[width=0.95\textwidth]{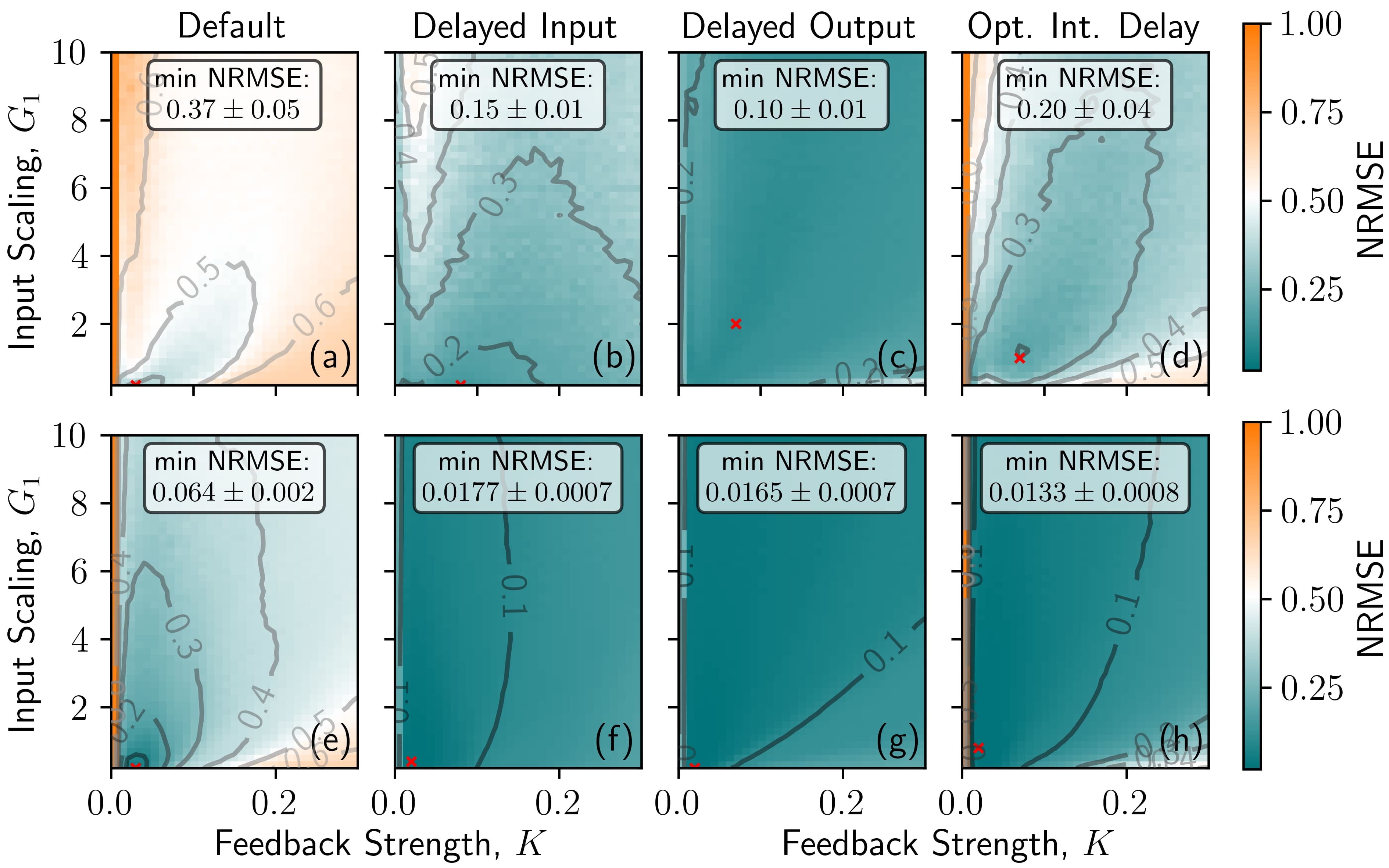}}
\caption{\textbf {Time-delayed reservoir} - NRMSE for the Mackey-Glass 12-step-ahead prediction task ($\Delta t =1$, $\Delta s=12$) color coded in dependence of the feedback strength $K$ and the input scaling $G_1$ for two reservoir sizes, $S=10$ (upper row (a-d) with $Z_R=50$) and $S=500$ (lower row (e-h) with $Z_R=10$) and 4 different delay schemes (columns): (b,f) delayed input with $G_2=0.6$, $n_I=8$, (c,g) delayed output with $n_S=8$, and (d,h) optimal internal delay $N=5S+1$. The minimum error for each subplot is given in the insets and the corresponding $G_1$-$K$ values are indicated by the red crosses. Remaining parameters as in Tab.~\ref{tab1}.}
\label{fig1}
\end{figure}

Figure~\ref{fig1}a shows the NRMSE for the Mackey-Glass 12-step-ahead prediction task as a function of the feedback strength $K$ and the input scaling $G_1$ for a default reservoir consisting of $S=10$ virtual nodes. The parameter ranges are chosen such that the characteristics of the nonlinearity span from linear to highly nonlinear and the feedback strength covers experimentally realisable values. Over the entire parameter plane, relatively poor performance is achieved (orange colors in Fig.\,\ref{fig1}a). Increasing the number of virtual nodes to $S=500$ (Fig.~\ref{fig1}e), the minimum NRMSE is decreased from $0.37\pm 0.05$ to $0.064\pm 0.002$. However, this comes at a cost of a factor 50 longer clockcycle $\tilde{T}=S$ and requires sensitive tuning of the reservoir parameters. 

To look at the influence of a delayed-input, as described in Sec.\ref{Sec:indel}, we select the parameters corresponding to the lowest error in Fig.~\ref{fig1}a (Fig.~\ref{fig1}e) and perform a grid search in the input-delay $n_I$ and input scaling $G_2$. Please see Fig.\,\ref{appfig1} in \ref{app1} for the corresponding scans. The $n_I$ and $G_2$ parameters for which the lowest error is achieved ($n_I=8,G_2=0.6$) are then used for the $G_1$-$K$ scan shown in Fig.~\ref{fig1}b (Fig.~\ref{fig1}f). The delayed-input parameters are therefore not optimised for each pair of values in the $G_1$-$K$ parameter plane, however, we have checked that the optimal $n_I$ does not vary significantly (see Fig.~\ref{appfig1b} in \ref{app1}). With the delayed input included, not only is the minimum error reduced, the variation of the error in dependence of $G_1$ and $K$ is also greatly reduced (mostly green colors in Fig.~\ref{fig1}f). 
For the delayed-state matrix concatenation method described in Sec.\ref{Sec:statemat}, we use the same approach as above, i.e. we optimise the state matrix delay $n_S$ for the optimal $G_1$,$K$ pair from Fig.~\ref{fig1}a (Fig.~\ref{fig1}e) and then use this  value, $n_S=8$, to produce Fig.~\ref{fig1}c (Fig.~\ref{fig1}g). The results of optimising the internal delay of the reservoir, again using the same approach as above, are shown in Fig.~\ref{fig1}d (Fig.~\ref{fig1}h). The optimised delay values for the Mackey-Glass 12 step ahead prediction task ($\Delta t=1$) are $n_I=8$, $n_S=8$ and an internal delay length of about 5 times the clock cycle $N=5S+1$.

\begin{figure}[t]
\centerline{\includegraphics[width=0.9\textwidth]{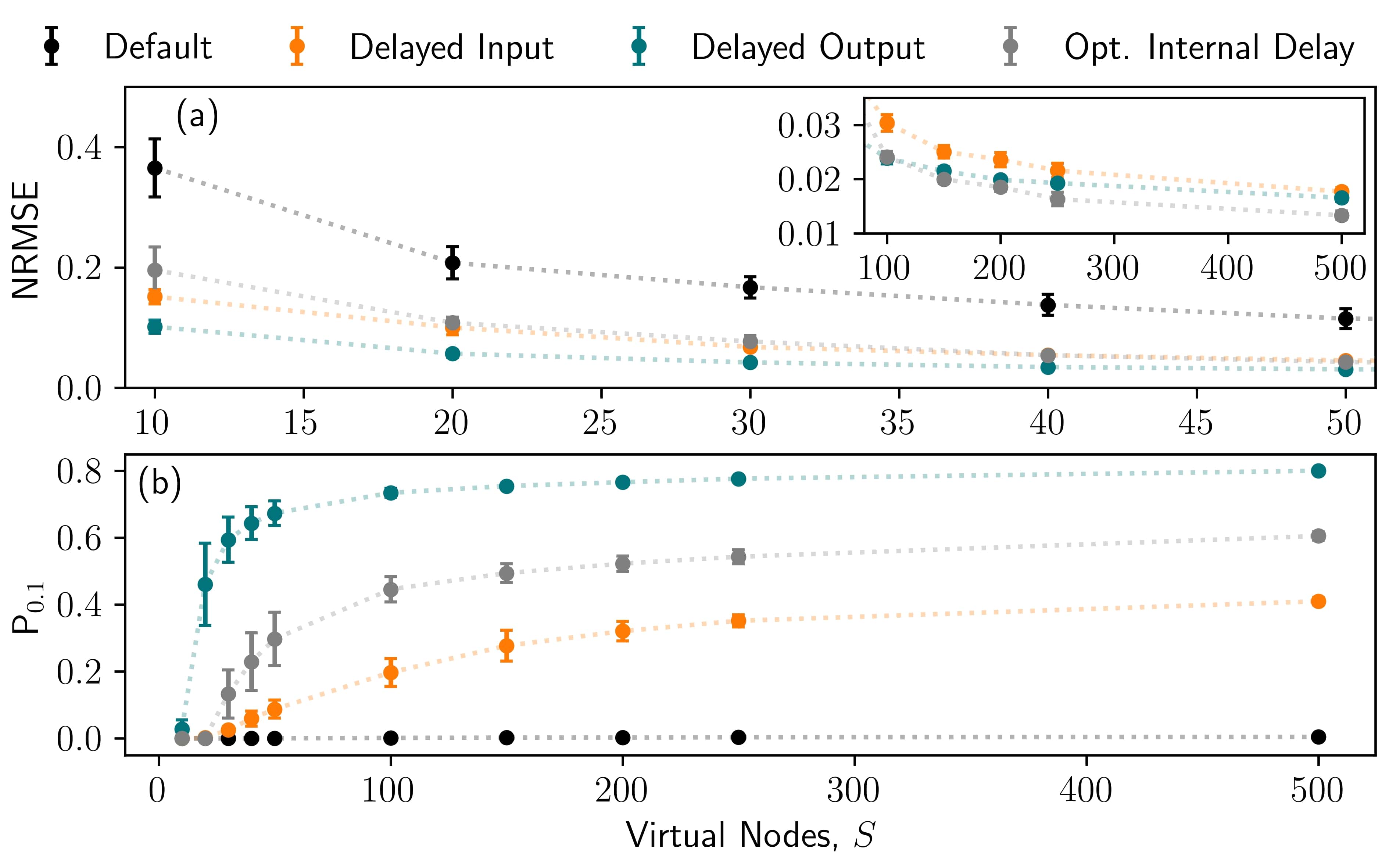}}
\caption{\textbf {Impact of reservoir size} - (a) Minimum NRMSE  and (b) parameter sensitivity $P_{0.1}$ for the Mackey-Glass 12-step-ahead prediction task ($\Delta t =1$, $\Delta s=12$) in dependence of the  virtual node number $S$ for the default delay based setup (black), with delayed input $G_2=0.6$ and $n_I=8$ (orange), with delayed output $n_S=8$ (green) and for optimal internal delay $N=5S+1$ (grey). Parameters: $Z_R=50$ for $S<100$ and as in Table~\ref{tab1}.}
\label{fig3}
\end{figure}

The colour distributions in the three rightmost columns of Fig.\,\ref{fig1}, show that all three methods of augmenting or changing the memory of the reservoir to optimise the performance for this Mackey-Glass task lead to a large decrease in the error over most of the depicted $G_1$-$K$ parameter plane (the minimum error achieved is always depicted in the inset of each panel). To summarise the results of Fig.~\ref{fig1}, and similar calculations for virtual node numbers between $S=10$ and $S=500$, we show the minimum error as a function of the number of virtual nodes $S$ in Fig.~\ref{fig3}a. To quantify the sensitivity of the absolute error to the reservoir parameters $G_1$ and $K$ we introduce a quantity $P_{0.1}$, which is the percentage of the depicted $G_1$-$K$ plane for which the NRMSE is under a threshold value of 0.1. Figure.~\ref{fig3}b shows $P_{0.1}$ as a function of the number of virtual nodes $S$ (results for a NRMSE threshold of 0.05 ($P_{0.05}$) are given in Fig.\ref{appfig3} in \ref{app2}). Figure~\ref{fig3} shows that, regardless of the method, including a task relevant time scale vastly improves the performance over a wide range of reservoir parameters. The smallest error in Fig.~\ref{fig3} is achieved when the internal delay $N$ of the reservoir is optimised (for $S=500$). Whereas, the least parameter sensitivity is achieved using the delayed state matrix concatenation method (dark green points in Fig.~\ref{fig3}b). Here we would like to note that there are several ways in which one could define a parameter sensitivity. We have chosen to quantify the sensitivity with respect to an absolute error, rather than, for example, with respect to the minimum error in each case. Our motivation for this is that for real world applications one would typically have an absolute threshold for what error is tolerable.
We do, however, show an example of the sensitivity defined relative to the respective minimum error in Fig.~\ref{appfig7} in \ref{app2}, and similar trends are found.

It should be noted that which of the memory augmentation methods produces the best results is strongly task dependent. This is because all methods lead to a different nonlinear mixing of the input signal.
An example where the delayed-input method out performs the delayed state matrix concatenation method in terms of the sensitivity measure $P_{0.1}$ is shown in Fig.\,\ref{appfig6}  in \ref{app2} where a Mackey Glass task one step prediction with $\Delta t =12$, $\Delta s=1$ was used. The most pronounced differences compared with Fig.~\ref{fig3} are seen for the default case with $N=S+1$. This is due to the larger discretisation $\Delta t =12$ enabling a much better delay embedding of the Mackey Glass timeseries and therefore the non-optimized internal delay configuration already leading to good prediction results. The next section will focus on this issue.


\subsection{Relationship between optimal delays, task and reservoir}\label{Sec:Resultsdiscretisation}

\begin{figure}[t]
\centerline{\includegraphics[width=0.95\textwidth]{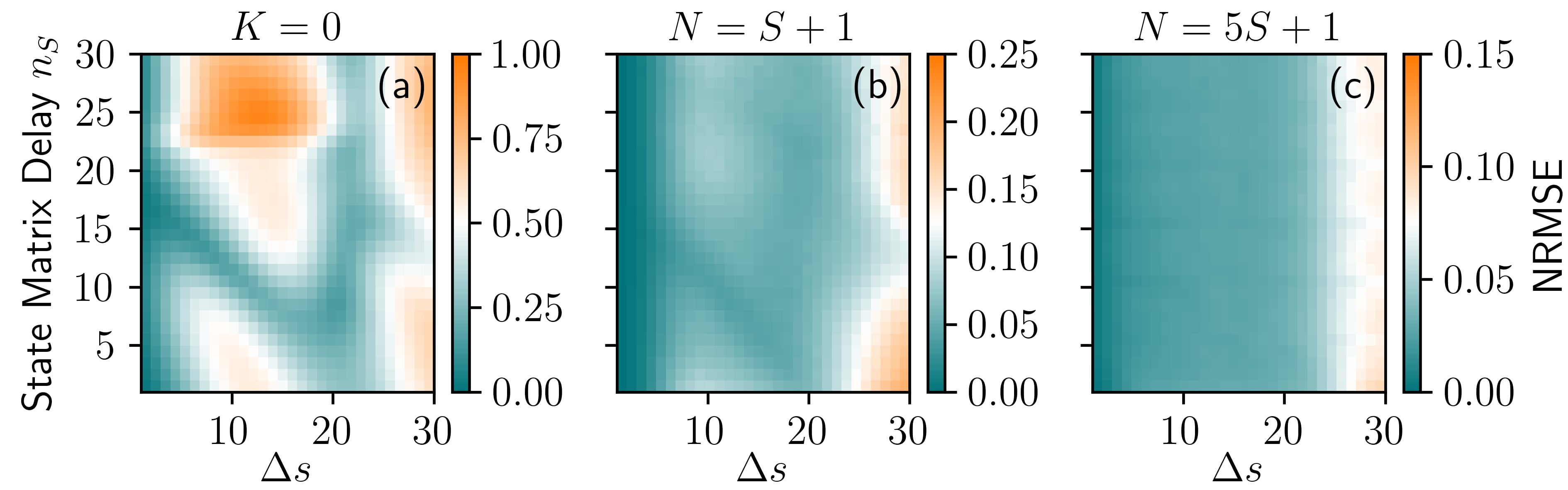}}
\caption{\textbf {Impact of prediction horizon on output delay} - NRMSE for the Mackey-Glass prediction task ($\Delta t=1$) color coded as a function of the prediction step $\Delta s$ and the externally added state matrix delay $n_S$. (a) extreme learning machine ($K=0$), (b) default internal delay setup, and (c) optimal internal delay setup. Parameters: $S=50$, $Z_R=50$ and remaining parameters as in Tab.~\ref{tab1}.}
\label{fig7}
\end{figure}

\begin{figure}[t]
\centerline{\includegraphics[width=0.95\textwidth]{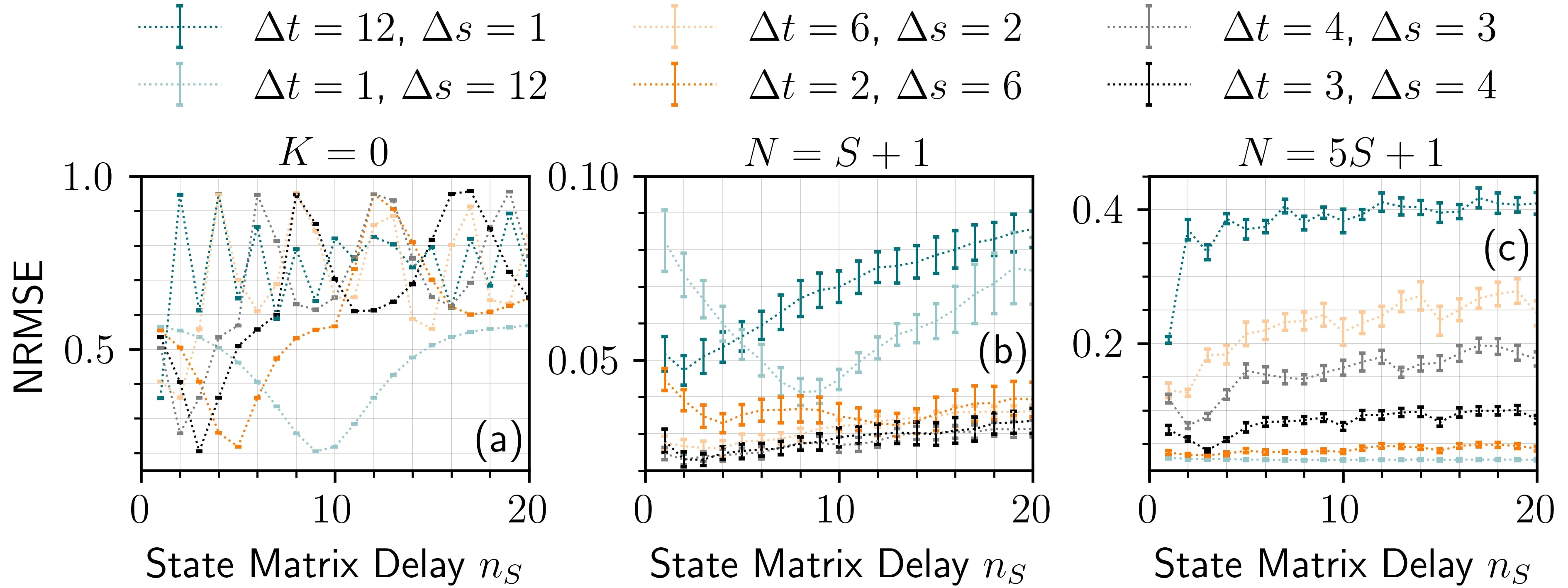}}
\caption{\textbf {Optimal output delay} - Impact of discretisation on the NRMSE for the Mackey-Glass prediction task for constant prediction horizon $\Delta s\cdot \Delta t$ as a function of the state matrix delay $n_S$. (a) extreme learning machine ($K=0$), (b) default internal delay of $S=N+1$, and (c) internal delay of $5S+1$.  Parameters: $S=50$, $Z_R=50$ and remaining parameters as in Tab.~\ref{tab1}.}
\label{fig8}
\end{figure}

The optimal augmentation delays ($n_I$ and $n_S$) depend on the requirements of the task and on the memory properties of the reservoir. We demonstrate this 
by altering the Mackey-Glass task. Firstly, by changing $\Delta s$ for a fixed discretisation $\Delta t$ (Fig.\,\ref{fig7}) and secondly, by altering $\Delta s$ and $\Delta t$ under the constraint that  $\Delta s \times \Delta t$ is fixed, i.e. that the time span of the prediction step is constant (Fig.\,\ref{fig8}). We then consider three implementations of the reservoir which have very different recall abilities. Results for these cases are shown in Fig.~\ref{fig7} and Fig.~\ref{fig8}. In Fig.~\ref{fig7}a and Fig.~\ref{fig8}a the feedback strength $K$ is set to zero. In this limit the reservoir itself is a single layer feedforward network, i.e. an extreme learning machine \cite{HUA04,ORT15}, and has no memory on its own. Figure~\ref{fig7}b and Fig.~\ref{fig8}b show results for the default configuration, where the internal delay is one time step longer than the input clockcycle. In this configuration the system behaves similar to a simple ring \cite{HUE22}; from one clockcycle to the next neighbouring nodes couple and the rate of information decay is determined by $K$. In Fig.~\ref{fig7}c and Fig.~\ref{fig8}c the internal delay is about five times as long as the input clockcycle, $N=5S+1$. This means that the reservoir has zero ability to recall information between one and four steps into the past. In all three cases the output delay $n_S$ of the state matrix concatenation adds additional memory.

For $K=0$, Fig.~\ref{fig7}a clearly shows that as $\Delta s$ is changed, so does the optimal input delay, ($n_s$ corresponding to the darkest green). The sum between optimal output delay and predicted step $\Delta s$, however stays constant until $\Delta s=19$ leading to a z-like behaviour of the optimal output delay. For the default internal delay configuration (Fig.~\ref{fig7}b), the optimal state matrix delay is shifted slightly to lower values compared with the $K=0$ case. This is because the state matrix delay compensates for the memory that is missing in the reservoir and for the default case there is memory due to the coupling to past reservoir states. When the internal delay is $N=5S+1$ (Fig.\,\ref{fig7}c) the reservoir is already well suited for delay embedding the Mackey-Glass input signal, therefore adding the state matrix concatenation has only a very small influence. Only slight improvements can be seen for $n_S$ values that are multiples of 5.

When the input discretisation $\Delta t$ is changed then the required memory can vary substantially, even if the prediction horizon $\Delta s\times \Delta t$ stays constant. This can be seen most clearly in Fig.~\ref{fig8}a by the varying position of the minimum error along the $n_S$ axis. However, $n_S$ is given in units of $\Delta t$ and the optimal value of $n_S$ in units of the original Mackey-Glass time series stays constant at about $n_S^{opt}\cdot \Delta t=9$ for the extreme learning limit $K=0$. For the default delay configuration  in Fig.\,\ref{fig8}b additional minima are seen.  Inspecting the minima of the curves in Fig.~\ref{fig8}b, it can also be seen that different input discretisations result in different lowest errors ($\Delta t=3$ and $\Delta t=4$ are the best choices here), which was not the case in the extreme learning limit in Fig.\,\ref{fig8}a. Thus, the optimal discretisation depends on the memory of the reservoir. 

In Fig.\,\ref{fig8}c, an internal delay configuration of $N=5S+1$ was chosen which was optimal for the discretisation of $\Delta t=1$ (cyan curve). If the discretisation is changed, $N$ is no longer optimized, and adding an output delay can now provide the missing memory. In general, a clear predictive relationship between the optimal augmentation delays, the task requirements and the properties of the reservoir are non-trivial. We thus recommend treating the delays as tuning parameters, guided by Takens embedding.

\subsection{Lorenz results}\label{Sec:Lresults}

So far we have used various versions of the Mackey-Glass timeseries prediction task. In this section we present a brief summary of results obtained using the Lorenz 63 system. We do this to make clear that the influence of the memory augmentation methods is not specific to the Mackey-Glass system and to show that the benefits of such methods are related to the memory requirements of the task. In Table~\ref{tabL} we show the minimum NRMSE, $P_{0.1}$ and $P_{0.05}$, which we have calculated using the same optimisation procedure as in Sec.~\ref{Sec:MGTDR}. For the $X(t)\rightarrow X\lb t+\Delta t\rb$ task the benefits of the memory augmentation methods are marginal. The state matrix concatenation method shows the best performance, both in terms of the absolute error and the parameter sensitivity. However, we remind the reader that the effective output dimension is doubled when the state matrix concatenation is used (see Sec.~\ref{Sec:statemat}).

\begin{table}
\caption{\label{tabL}\textbf {Time-delayed reservoir} - Lorenz 63 timeseries prediction performance for 3 different tasks (rows) and 3 different delay configurations (columns). Parameters as in Tab.~\ref{tab1}.}
\begin{indented}
\item[]\begin{tabular}{@{}llll}
\br
&\multicolumn{3}{l}{\hspace{-1cm}Task: Lorenz X, one step ahead with $\Delta t$=0.1}\\
&\multicolumn{3}{l}{$X(t) \rightarrow X(t+\Delta t)$}\\
\mr
\multicolumn{1}{c|}{}&Default& Delayed Input & Delayed Output \\
\multicolumn{1}{c|}{}&& $n_I=2$, $G_2=0.2$ & $n_s=1$ \\
\mr
\multicolumn{1}{c|}{min NRMSE}&$0.0193\pm 0.0007$&$0.0176\pm0.0007$&${\bf 0.0140\pm 0.0006}$\\
\multicolumn{1}{c|}{$P_{0.1}$}&$0.790\pm 0.009$&$0.830\pm 0.006$&${\bf 0.882\pm 0.003}$\\
\multicolumn{1}{c|}{$P_{0.05}$}&$0.47\pm 0.01$&$0.53\pm0.01$&${\bf 0.597\pm0.008}$\\
\br
&\multicolumn{3}{l}{\hspace{-1cm}Task: Lorenz cross prediction with $\Delta t$=0.1}\\
&\multicolumn{3}{l}{$X(t) \rightarrow Z(t)$}\\
\mr
\multicolumn{1}{c|}{}&Default& Delayed Input & Delayed Output \\
\multicolumn{1}{c|}{}&& $n_I=2$, $G_2=0.2$ & $n_s=2$ \\
\mr
\multicolumn{1}{c|}{min NRMSE}&$0.0058\pm 0.0009$&$0.0046\pm0.0004$&${\bf 0.0030\pm 0.0002}$\\
\multicolumn{1}{c|}{$P_{0.1}$}&$0.41\pm 0.02$&$0.54\pm 0.02$&${\bf 0.898\pm 0.008}$\\
\multicolumn{1}{c|}{$P_{0.05}$}&$0.13\pm 0.02$&$0.24\pm0.03$&${\bf 0.55\pm0.01}$\\
\br
&\multicolumn{3}{l}{\hspace{-1cm}Task: Lorenz cross prediction with $\Delta t$=0.02}\\
&\multicolumn{3}{l}{$X(t) \rightarrow Z(t)$}\\
\mr
\multicolumn{1}{c|}{}&Default& Delayed Input & Delayed Output \\
\multicolumn{1}{c|}{}&& $n_I=17$, $G_2=0.2$ & $n_s=20$ \\
\mr
\multicolumn{1}{c|}{min NRMSE}&$0.099\pm 0.006$&$0.030\pm0.002$&${\bf 0.0167\pm 0.0005}$\\
\multicolumn{1}{c|}{$P_{0.1}$}&$0.0003\pm 0.0003$&${\bf 0.09\pm 0.02}$&$0.044\pm 0.006$\\
\multicolumn{1}{c|}{$P_{0.05}$}&$0\pm 0$&$0.007\pm0.001$&${\bf 0.012\pm0.001}$\\
\br
\end{tabular}
\end{indented}
\end{table}

For the cross-prediction task, $X\lb t\rb\rightarrow Z\lb t\rb$, the memory augmentation methods greatly improve the results when $\Delta t=0.02$, but lead to only a slight improvement for $\Delta t=0.1$. However, better performance is achieved for the $\Delta t=0.1$ case. This is because $\Delta t=0.1$ is the optimal delay for delay embedding the Lorenz 63 system \cite{STO22,KAN03}. This means that with $\Delta t=0.02$ the reservoir needs to retain information from further in the past and, hence, augmenting the memory has a greater impact.

Although not shown here, in contrast to the cross-prediction task, the performance for the  $X(t)\rightarrow X\lb t+\Delta t\rb$ task improves when $\Delta t$ is smaller than 0.1. This can be understood by looking at the autocorrelation plot in Fig.~\ref{corrLor}c. For small lags the autocorrelation of $X\lb t\rb$ is high, whereas the cross-correlation with $Z \lb t\rb$ is close to zero even for small lags. This means that the $X(t)\rightarrow X\lb t+\Delta t\rb$ task becomes more linear and therefore easier as $\Delta t$ approaches zero, however, for the $X\lb t\rb\rightarrow Z\lb t\rb$ task this is not the case.

\subsection{Random Network}

\begin{figure}[b]
\centerline{\includegraphics[width=0.9\textwidth]{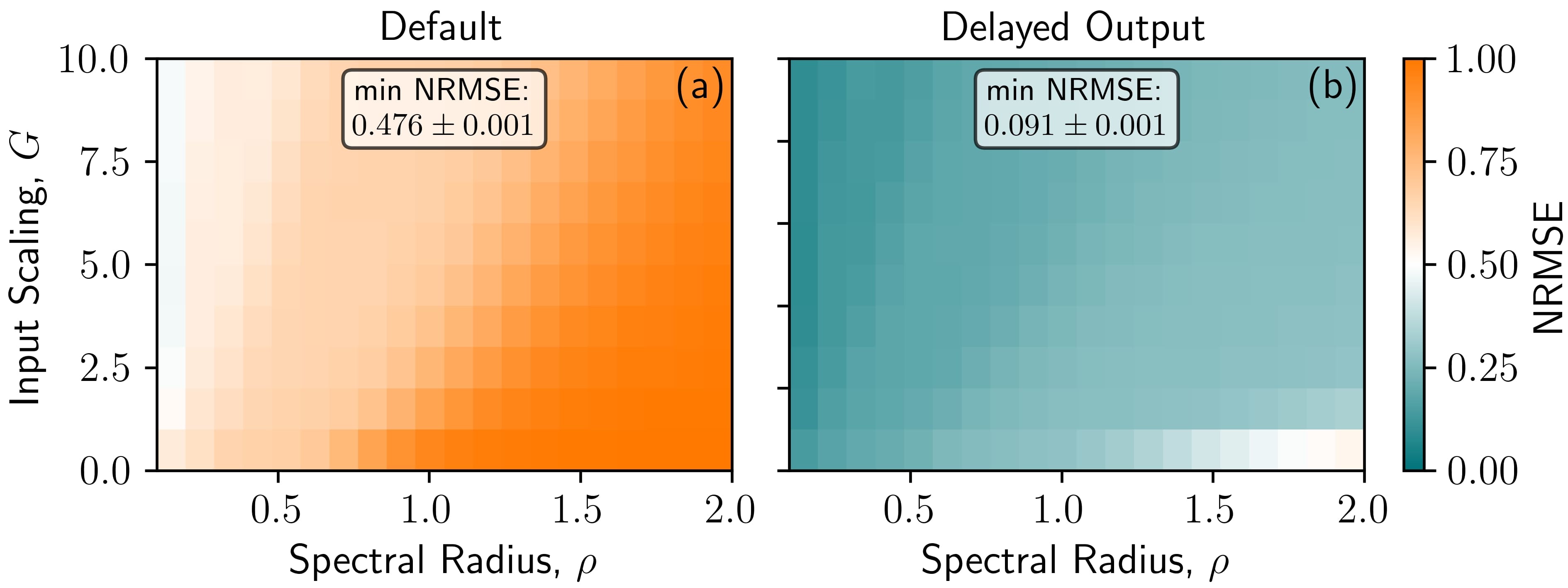}}
\caption{\textbf {Random network} - NRMSE for the Mackey-Glass 12-step-ahead prediction task in dependence of the spectral radius $\rho$ and the input scaling $G_1$. Parameters: $n_S=8$, $S=500$ and $Z_R=10$.}
\label{figrand}
\end{figure}

Delay based reservoirs and the more commonly used ESNs, or random recurrent networks, have very different memory properties. In the case of the ESN the memory is determined indirectly by the random topology of the reservoir and the spectral radius, and will mostly be monotonically decreasing \cite{LYM19a}. However, for a delay-based reservoir, such as the one used here, the memory can be directly tuned by changing the ratio of the delay and the input clocktime, and will not decrease monotonically if the delay is sufficiently large compared with the clocktime and the memory of the solitary nonlinear node \cite{ROE19,KOE21,HUE22}. These differences, however, have no bearing on the influence of the memory augmentation methods. We demonstrate this in Fig.~\ref{figrand} which shows one example of the performance improvement that can be achieved with the delayed state matrix concatenation method. Compared with the time-delayed reservoir (see Fig.~\ref{fig1}), the overall performance is worse because our chosen nonlinearity produces a binary output (Sec.~\ref{Sec:modelrand}). We also note, that for Fig.~\ref{figrand} we did not optimise the delay $n_S$, we simply set it to $n_S=8$, it is therefore possible that the results could be improved further. However, despite this a stark improvement is obtained over the entire parameter range when the delayed state matrix concatenation is included.

\section{Conclusion}\label{Sec:Conclusion}

We have shown that including task specific timescales can significantly improve the performance of a reservoir computer for timeseries prediction tasks. While the timescale itself matters, the method of including it does not play such an important role for the overall performance and should be chosen according to the requirements of the specific task.
Our focus was on a simple means of improving the performance, which reduces the need for hyperparameter tuning in the context of hardware implementation. There are advantages and disadvantages to all methods. The state matrix concatenation method has the advantage that it does not influence the reservoir as it comes after the reservoir sampling step. This is likely the reason why this method showed the best performance in terms of the hyperparameter sensitivity. It also has the advantage that it only introduces one tuning parameter, the delay. However, if this method is to be implemented in hardware, the previous reservoir states need to be stored and the state vector is twice as large. The delayed input method, on the other hand, has the comparative disadvantage that it introduces two tuning parameters, the delay and the input scaling. However, this method produces very similar results to the state matrix concatenation without an increased effective output dimension. In the case of a delay-based reservoir, tuning the internal delay is also an option, but can be difficult in practise, particularly in photonic systems which can be very feedback phase sensitivity \cite{GLO12,LIN19,LUE20}. Lastly, one can optimise the discretisation of the input signal, but here one must also consider the desired output discretisation.




Extensions or variations of the methods discussed here are possible and some have already been study \cite{TAK18,MAR19a,SAK20,PAU21,DEL21a,CAR22a}. Whether or not to apply such methods once again comes down to the objective and a trade-off between simplicity and absolute performance. 




\appendix

\section{Delay optimisation for augmented memory}\label{app1}
\begin{figure}[t]
\centerline{\includegraphics[width=0.95\textwidth]{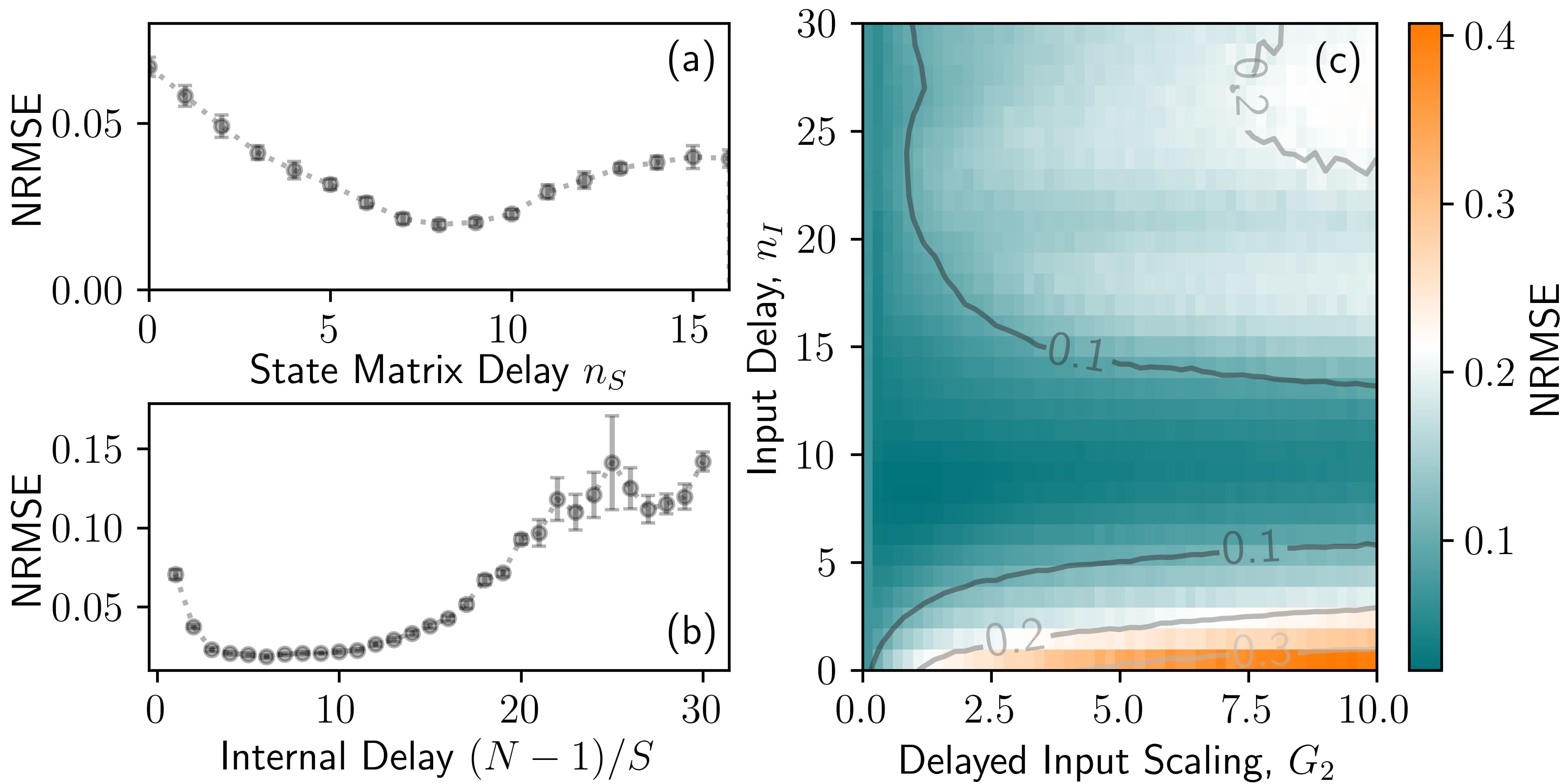}}
\caption{\textbf {Optimizing delays} - NRMSE for the Mackey-Glass 12-step-ahead prediction task ($\Delta t=1$) in dependence of (a) state matrix concatenation delay $n_S$, (b) internal delay $N$ (normalized to the number of virtual nodes $S$ and chosen off-resonant), and (c) input delay $n_I$ and input scaling $G_2$. For parameters see Tab.~\ref{tab1}.}
\label{appfig1}
\end{figure}

Figure~\ref{appfig1} shows the grid search and parameters scans performed to find the optimal internal delay $N$, the optimal input delay $n_I$ and the best state matrix concatenation delay $n_S$. For the internal delay scan $N$ was increased in steps of $S$ starting at $N=1$. This was done to avoid resonances between the delay $N$ and clock cycle $\tilde{T}$, which are known to be detrimental \cite{JAU21a,STE20}.

\begin{figure}[t]
\centerline{\includegraphics[width=0.95\textwidth]{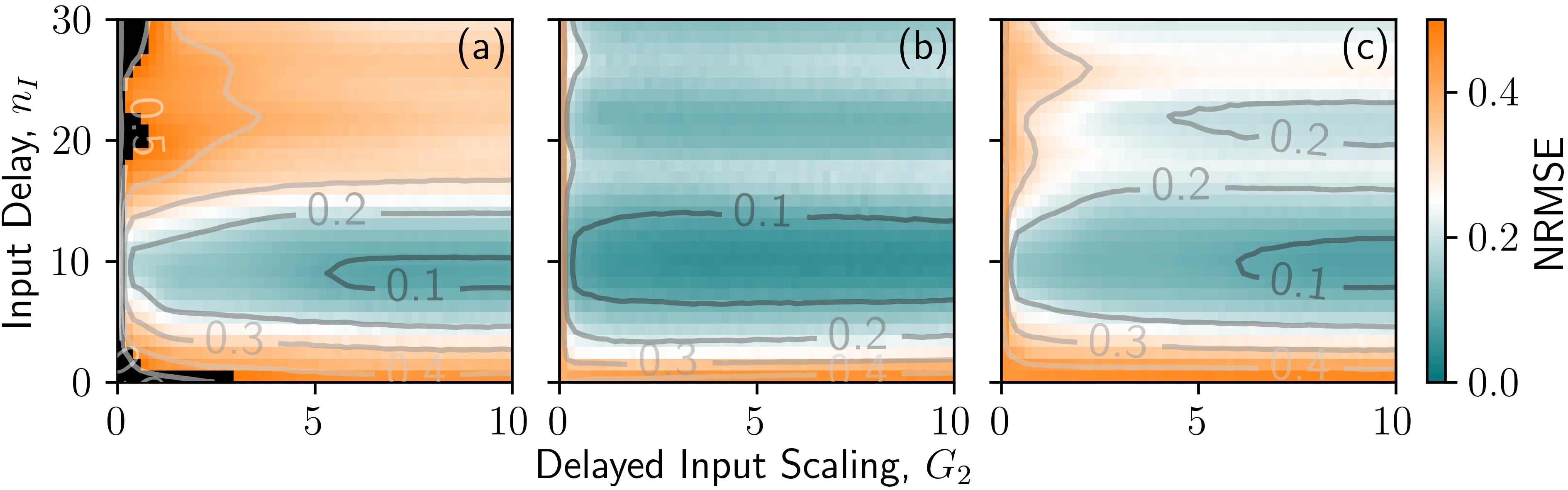}}
\caption{\textbf {Optimizing delays} - NRMSE for the Mackey-Glass 12-step-ahead prediction task ($\Delta t=1$) in dependence of the input delay $n_I$ and input scaling $G_2$ for (a) $K=0.3$, $G_1=0.2$, (b) $K=0.03$, $G_1=10$ and (c) $K=0.3$, $G_1=10$. Remaining parameters as in Tab.~\ref{tab1}.}
\label{appfig1b}
\end{figure}

In Fig.~\ref{appfig1b} examples of input delay scans are shown for three additional $G_1$-$K$ pairs. For these three pairs of $G_1$-$K$ values the nonlinearity and memory properties are very different, as is the performance of the default reservoir (see Fig.~\ref{fig1}a,e). However, in all cases the optimal input delay $n_I$ is in the range from 8 to 12 and the reservoirs all perform well within this range.

\section{Further Mackey-Glass Performance Results}\label{app2}

\subsection{Parameter sensitivity}
\begin{figure}[t]
\centerline{\includegraphics[width=0.9\textwidth]{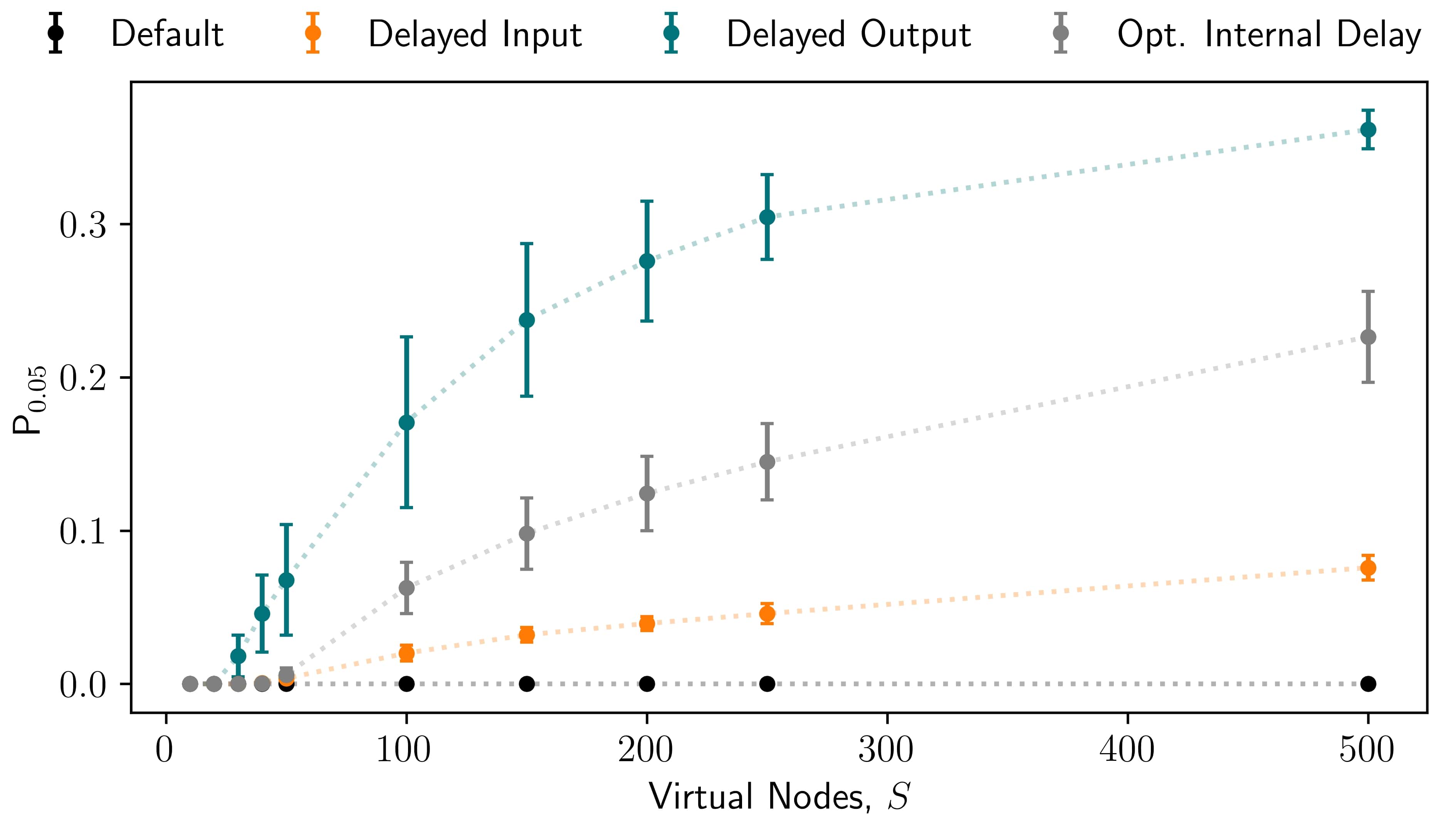}}
\caption{\textbf {Time-delayed reservoir} - Sensitivity measure $P_{0.05}$ for the Mackey-Glass 12-step-ahead prediction task $\Delta t=1$ in dependence of the number of virtual nodes $S$, obtained from the results depicted in Fig.~\ref{fig1}. Parameters:  $N=S+1$ (black), $n_I=8$ and $G_2=0.6$(orange),  $n_S=8$ (green), $N=5S+1$ (grey), $Z_R=50$ for $S<100$ and as in Table~\ref{tab1}.}
\label{appfig3}
\end{figure}

Figure~\ref{appfig3} shows the parameter sensitivity measure $P_{0.05}$ for the Mackey-Glass 12 step-ahead prediction task with $\Delta t=1$. $P_{0.05}$ is the percentage of the $G_1$-$K$ parameter space which is depicted in Fig.~\ref{fig1} for which the NRMSE is below 0.05. 

\begin{figure}[t]
\centerline{\includegraphics[width=0.9\textwidth]{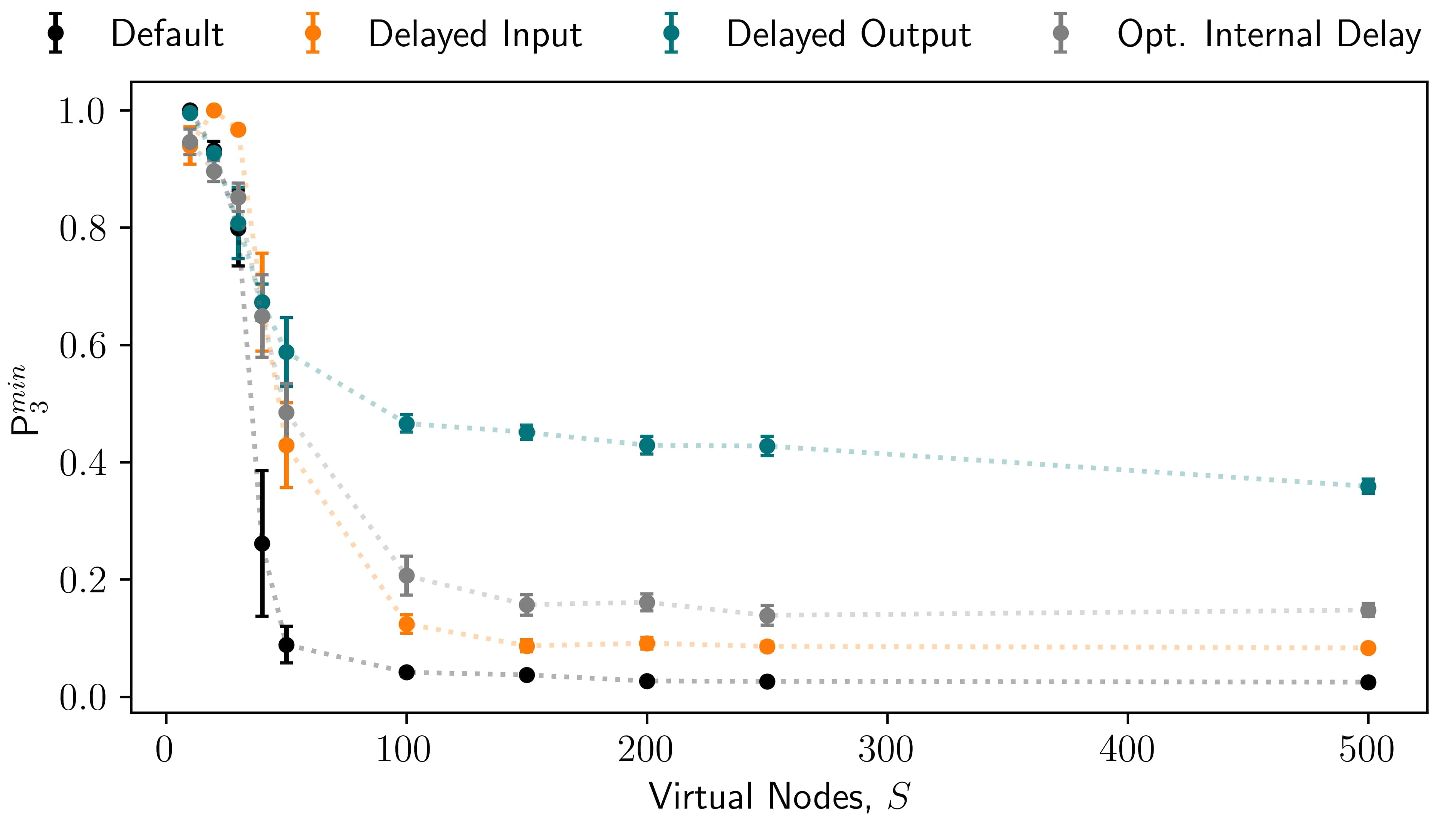}}
\caption{\textbf {Time-delayed reservoir} - Sensitivity measure $P^{min}_{3}$ for the Mackey-Glass 12-step-ahead prediction task $\Delta t=1$ in dependence of the number of virtual nodes $S$, obtained from the results depicted in Fig.~\ref{fig1}. Parameters:  $N=S+1$ (black), $n_I=8$ and $G_2=0.6$(orange),  $n_S=8$ (green), $N=5S+1$ (grey), $Z_R=50$ for $S<100$ and as in Table~\ref{tab1}.}
\label{appfig7}
\end{figure}

The parameter sensitivity can also be defined with respect to the minimum error achieved in each of the setups. An example of this is shown in Fig.~\ref{appfig7}. Here we have defined the sensitivity measure $P^{min}_{3}$ as the percentage of the parameter space for which the error is below three times the minimum error achieved for that reservoir setup. For node numbers greater than 50 Fig.~\ref{appfig7} shows the same trends as Fig.~\ref{fig3} and Fig.~\ref{appfig3}. However, this measure fails to provide useful information when the absolute error is very large, i.e. when the performance is uniformly poor, as is the case for low number of virtual nodes in the default reservoir case.

\subsection{Input discretisation}

\begin{figure}[t]
\centerline{\includegraphics[width=0.9\textwidth]{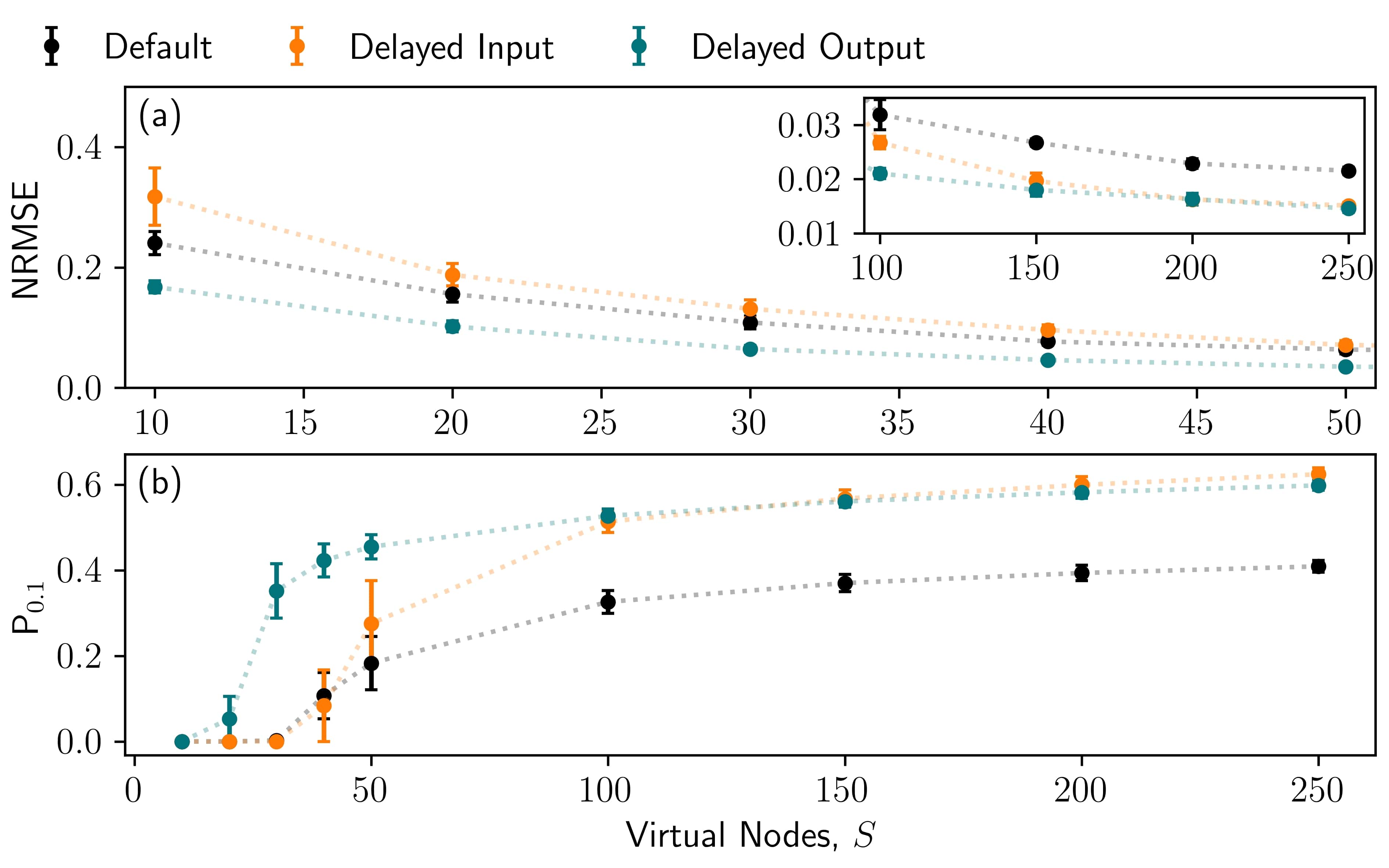}}
\caption{\textbf {Time-delayed reservoir} - (a) Minimum NRMSE  and (b) sensitivity measure $P_{0.1}$ for the Mackey-Glass $\Delta s=1$ step-ahead prediction task with $\Delta t =12$ in dependence of the number of virtual nodes $S$ for the default delay based setup (black), with delayed input $G_2=0.6$ and $n_I=8$ (orange), and with delayed output $n_S=8$ (green). Parameters: $Z_R=50$ for $S<100$ and as in Table~\ref{tab1}.}
\label{fig5}
\end{figure}

\begin{figure}[t]
\centerline{\includegraphics[width=0.9\textwidth]{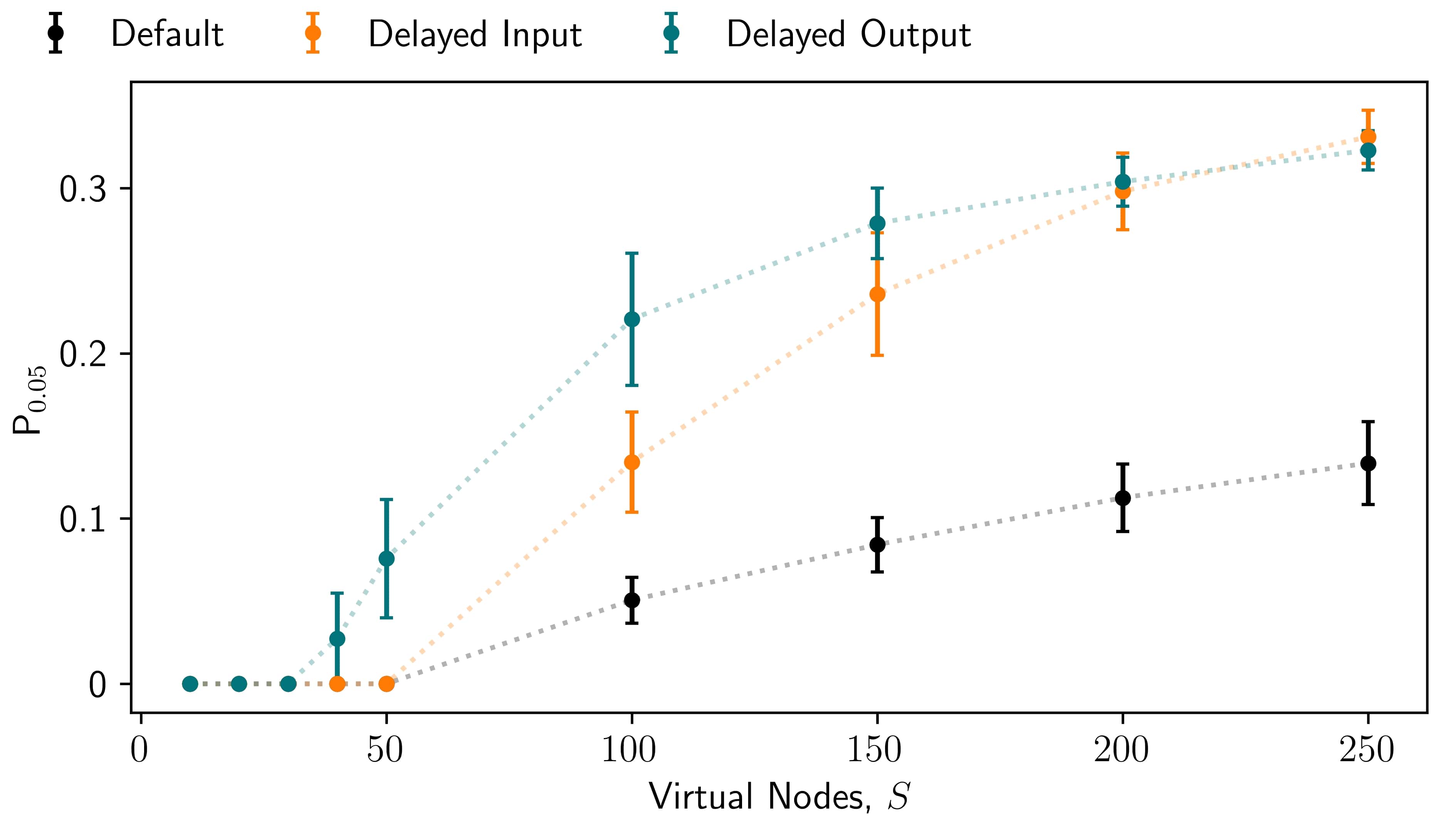}}
\caption{\textbf {Impact of reservoir size} - Sensitivity measure $P_{0.05}$ for the Mackey-Glass $\Delta s=1$ step-ahead prediction task with $\Delta t =12$ in dependence of the number of virtual nodes $S$ for the default delay based setup (black), with delayed input $G_2=0.6$ and $n_I=8$ (orange), and with delayed output $n_S=8$ (green). Parameters: $Z_R=50$ for $S<100$ and as in Table~\ref{tab1}.}
\label{appfig6}
\end{figure}

Analogous to Figs~\ref{fig3}, Fig.~\ref{fig5} shows the minimum NRMSE and the sensitivity $P_{0.1}$ for an input discretisation of $\Delta t=12$ and one step-ahead prediction ($\Delta s=1$), i.e. the size of the prediction step is the same as in the $\Delta t =1$ and $\Delta s =12$ case. The default reservoir now shows much better performance. This is because less redundant information is being feed into the reservoir and the memory requirements are reduced. The performance is still improved when a delay-input or delayed state matrix concatenation is included, however the impact is reduced and the optimised delays are smaller (now $n_I=4$ and $n_S=4$, compared with $n_I=8$ and $n_S=8$ for the  $\Delta t =1$ and $\Delta s =12$ case). We also note that, in terms of the sensitivity measure $P_{0.1}$, the delayed input method out performs the delayed state matrix concatenation method for large node numbers, even though the effective state matrix dimension is only half as big. This also holds for $P_{0.05}$, see Fig.~\ref{appfig6}.

\clearpage

\section*{References}
\bibliographystyle{iopart-num}
\providecommand{\newblock}{}

\end{document}